\documentstyle[12pt,epsfig]{article}

\newcommand{\beq}{\begin{equation}}
\newcommand{\eeq}{\end{equation}}
\newcommand{\bea}{\begin{eqnarray}}
\newcommand{\eea}{\end{eqnarray}}
\newcommand{\nn}{\nonumber}

\newcommand{\gtrsim}{\ \rlap{\raise 2pt\hbox{$>$}}{\lower 2pt \hbox{$\sim$}}\ }
\newcommand{\lessim}{\ \rlap{\raise 2pt\hbox{$<$}}{\lower 2pt \hbox{$\sim$}}\ }

\newcommand{\np}[1]{Nucl. Phys. {\bf #1}}
\newcommand{\pl}[1]{Phys. Lett. {\bf #1}}
\newcommand{\pr}[1]{Phys. Rev. {\bf #1}}
\newcommand{\prl}[1]{Phys. Rev. Lett. {\bf #1}}

\newcommand{\rmp}[1]{Rev. Mod. Phys. {\bf #1}}    
\newcommand{\ijmp}[1]{Int. Jour. Mod. Phys. {\bf #1}}

\makeatletter
\if@twoside
\else \oddsidemargin 00pt \evensidemargin 00pt
\fi
\let\@eqnsel = \hfil

\expandafter\ifx\csname mathrm\endcsname\relax\def\mathrm#1{{\rm #1}}\fi
\@ifundefined{mathrm}{\def\mathrm#1{{\rm #1}}}{\relax}

\makeatother

\begin{document}

\title{\vskip-2.5truecm{\hfill \baselineskip 14pt {{
\small  \\
\hfill MZ-TH/98-29 \\ 
\hfill July  1998}}\vskip .9truecm}
 {\bf Only Three}}

\vspace{5cm}

\author{Gabriela Barenboim\footnote{\tt gabriela@thep.physik.uni-mainz.de} 
\phantom{.}and Florian Scheck\footnote{\tt Scheck@dipmza.physik.uni-mainz.de}
 \\  \  \\
{\it  Institut f\H ur Physik - Theoretische 
Elementarteilchenphysik }\\
{\it Johannes Gutenberg-Universit\H at, D-55099 Mainz, Germany}
\\
}

\date{}
\maketitle
\vfill

\begin{abstract}
\baselineskip 20pt
It is shown that it is possible to account for all the
experimental indications for neutrino oscillations with just
three flavours.
In particular we suggest that the atmospheric neutrino anomaly and
the LSND result can be explained by the same mass difference 
and mixing. Possible implications and future test of the resulting mass
and mixing pattern are given.

\end{abstract}
\vfill
\thispagestyle{empty}

\newpage
\pagestyle{plain}
\setcounter{page}{1}

\section{Introduction} 

Currently there are three pieces of evidence which suggest that
neutrinos may
have non zero mass differences and nontrivial mixings. These are the anomalies
observed in the solar neutrino flux \cite{re1}, in atmospheric neutrino
production \cite{re2,re2b}  
and in neutrino beams from accelerators and reactors
\cite{re3}.
The conventional approach in analyzing any single experiment 
consists in parametrizing the neutrino 
anomaly in terms of oscillations between two
neutrino mass states only. Clearly, this assumption is too restrictive
when one sets out to fit several anomalies simultaneously and a consistent
mixing scheme involving at least three lepton generations is
mandatory \cite{re}. In fact, a two-generation parametrization of a 
given experimental
result may lead to erroneous conclusions about the magnitude of the squared
mass differences and/or mixing matrix elements.

In this note we address the question whether all extent observations
can be explained simultaneously in terms of three neutrino mass
eigenstates which mix and oscillate. In contrast to earlier fits which 
seemed to suggest that one needed at least four generations \cite{bilenky}
we show that one can accommodate all experimental findings in a
setting with just three flavours. We find essentially two solutions
that describe all data and both of which are characterized by large
mixing between the three generations. We also show that the need for a 
fourth generation becomes obsolete because of a significantly
different choice of the relevant mass differences. 
The mixing pattern predicted by these solutions can be subject to
further experimental tests and, hence, may be verified or
disproven in ongoing or future experiments.

Regarding the mass spectrum of the three neutrinos we assume that it
is characterized by two differences of squared masses
\[
\Delta m^2\equiv \Delta m_{21}^2 = m^2_2 -m^2_1\quad\mbox{and } 
\Delta M^2\equiv\Delta m_{32}^2 = m^2_3 -m^2_2
\]
whose numerical values are rather different. The first of these 
is chosen in view of the solar neutrino  deficit, 
\bea
10^{-4}\, eV^2 \le \Delta m^2 \le 10^{-3}\, eV^2\, .
\eea
The lower limit is suggested by the results of Super KAMIOKANDE where
a significant effect would be seen only for values at this level or
higher. The upper limit is obtained from the CHOOZ reactor experiment
\cite{re35}
which showed that $\Delta m^2\le 10^{-3} eV^2$.

The second is selected with both the LSND and the atmospheric
neutrino data in mind, viz.
\bea
\Delta M^2 \simeq .3\; eV^2
\eea
This choice reflects the values allowed by the
LSND experiment, adding to them the constraint obtained from the 
Bugey reactor \cite{re4}, 
$\Delta M^2 > .2\; eV^2$, and the upper limit obtained in 
searches for oscillations by CDHSW  
\cite{re5} 
which gave $\Delta M^2 < .4\; eV^2$.

With these assumptions on the mass differences the LSND, the solar and
the atmospheric neutrino data can be explained
by an appropriate choice of the mixing matrix that defines the weak
interaction states.

The neutrino survival and transition probabilities are given by
\bea
P(\alpha \rightarrow \beta) = \delta_{\alpha \beta} -
 4 \sum_{i>j=1 }^3 U_{\alpha i} U_{\beta i} U_{\alpha j} U_{\beta j}
\sin^2 \left[ \frac{\Delta m_{ij}^2 L}{4 E} \right]
\eea
Here the matrix $U=\left\{ U_{\alpha i}\right\}$ describes the weak
interaction states ($\nu_\alpha$) in terms of the mass eigenstates
($\nu_i$). That is,
\bea
\nu_\alpha = \sum_i U_{\alpha i} \nu_i .
\eea
For the time being we ignore possible CP violation but return to this 
point later. As a consequence $U$ is taken to be real. It is
parametrized as follows
\bea
U=\pmatrix{c_1c_3 & s_1c_3 & s_3 \cr
       -s_1c_2 - c_1s_2s_3 & c_1c_2- s_1s_2s_3 & s_2c_3 \cr
        s_1s_2 - c_1c_2s_3 & - c_1s_2 - s_1c_2s_3 &c_2c_3}
\eea
In Eq. (3) $L$ denotes the neutrino flight path, i.e. the distance 
between the neutrino source and the detector,
$E$ is the energy of the neutrino in the laboratory system.

\section{The LSND experiment}

The Liquid Scintillation Neutrino Detector (LSND) experiment at Los Alamos
reported to have observed appearance of $\bar{\nu}_e$ in an
initial beam of $\bar{\nu}_\mu$'s \cite{re4}. 
These results were interpreted as
evidence for neutrino oscillations.  

The probability to observe oscillations is given by
\bea
P_{\mbox{LSND}} &= &-4 U_{23} U_{13} U_{21} U_{11} 
\sin^2 \left[ \frac{(\Delta m^2 + \Delta M^2) L}{4 E} \right] \nn \\
&& -4 U_{23} U_{13} U_{22} U_{12} 
\sin^2 \left[ \frac{\Delta M^2 L}{4 E} \right] 
-4 U_{22} U_{12} U_{21} U_{11} 
\sin^2 \left[ \frac{\Delta m^2  L}{4 E} \right]
\label{lsnd}
\eea
The LSND set up \cite{setup}
is characterized by $L$=30 m and $ 36 MeV < E < 60 MeV$.
With our  choice  of $\Delta m^2$, this means that the third term in 
Eq.(\ref{lsnd}) and the contribution of $\Delta m^2$ in the first term 
can be neglected. Combining then the first two terms and making
use of the unitarity of $U$ we have
\bea
P_{\mbox{LSND}} &\approx & 4 U_{23}^2 U_{13}^2  
\sin^2 \left[ \frac{\Delta M^2 L}{4 E} \right] \nn \\
&=& 4 s_2^2 c_3^2 s_3^2 \sin^2 \left[ \frac{\Delta M^2 L}{4 E} \right]
\eea

Hence, the requirement that the LSND results 
$P_{\mbox{LSND}}=.0031 \pm .0011 (\mbox{stat}) \pm .0005 (\mbox{syst})$
be consistent with our choice of parameters requires
\bea
s_2 c_3 s_3 = .104\, ,
\label{uno}
\eea
where we have taken the average energy ($E$=42 MeV).

Note that if in Eq. (7) one had inserted a very {\em small\/} squared mass
difference it would have been impossible to meet the requirement
(\ref{uno}). This may have been the reason that led the authors of
\cite{bilenky}  to conclude that a fourth generation was necessary to
explain the Los Alamos data.

\section{Atmospheric neutrinos}
The evidence for an anomaly in atmospheric neutrino experiments was pointed
out by the Kamiokande collaboration and by the IMB Collaboration \cite{re55}. 
More recently, the Super Kamiokande collaboration reported further and
more precise results on the anomaly in atmospheric neutrino fluxes
\cite{re2b}.

Oscillations of muon neutrinos into tau neutrinos as presumably
observed by Super Kamiokande are given by
\bea
P_{\mbox{Super K}} &= &-4 U_{23} U_{33} U_{21} U_{31} 
\sin^2 \left[ \frac{(\Delta m^2 + \Delta M^2) L}{4 E} \right] \nn \\
&& -4 U_{23} U_{33} U_{22} U_{32} 
\sin^2 \left[ \frac{\Delta M^2 L}{4 E} \right] 
-4 U_{22} U_{32} U_{21} U_{31} 
\sin^2 \left[ \frac{\Delta m^2  L}{4 E} \right]
\eea
Again, with our choice of squared mass differences, the contributions
of
$\Delta m^2$ can be neglected and the first two terms can be
combined into a single one, yielding
\bea
P_{\mbox{Super K}}(\nu_\mu\to\nu_\tau ) &\approx & 4 U_{23}^2 U_{33}^2  
\sin^2 \left[ \frac{\Delta M^2 L}{4 E} \right] 
\eea
Thus by averaging the sine squared factor to one half (which is
consistent with the variation in $E$ and $L$), we obtain,
\bea
P_{\mbox{Super K}}= 2 s_2^2 c_2^2 c_3^4 
\label{superk}
\eea

There are good reasons to believe that the overall deficit 
of muon neutrinos observed in Super Kamiokande atmospheric data for 
zenith angles whith $\cos \theta_z>-.6$ is primarily an effect
of muon neutrinos oscillating into tau neutrinos. Indeed, this is
supported by  
the smallness of the LSND effect and the non-observation of electron
neutrino disappearance in reactor experiments.
However, as we are working in a setting of three neutrino generations,
oscillations of muon neutrinos into electron neutrinos are not only 
possible but also mandatory in our scheme. Super Kamiokande 
should observe them with a probability given by
\bea
P_{\mbox{Super K}}(\nu_\mu\to\nu_e) &= &-4 U_{23} U_{13} U_{21} U_{11} 
\sin^2 \left[ \frac{(\Delta m^2 + \Delta M^2) L}{4 E} \right] \nn \\
&& -4 U_{23} U_{13} U_{22} U_{12} 
\sin^2 \left[ \frac{\Delta M^2 L}{4 E} \right] 
-4 U_{22} U_{12} U_{21} U_{11} 
\sin^2 \left[ \frac{\Delta m^2  L}{4 E} \right]
\eea
Again, the terms in $\Delta m^2$ are negligible, the first two terms
can be 
combined in a single one and their sine squared
factors can be averaged to one half, resulting in
\bea
P_{\mbox{Super K}}(\nu_\mu\to\nu_e) &\simeq & 2 U_{23}^2 U_{13}^2 -
4 U_{22} U_{12} U_{21} U_{11}   
\sin^2 \left[ \frac{\Delta m^2 L}{4 E} \right] 
\eea
If the mixing angles $\theta_2$ and $\theta_3$ are small, this implies 
that, for most of the zenith angle range, the atmospheric neutrino 
anomaly is indeed due to muon neutrinos oscillating into tau
neutrinos. We will see below that our fit does indeed favor small
values for these angles. 
Nevertheless, the second term will also be observable in upward 
going events provided $\Delta m^2 L/(4 E) $ is larger than or of order one.

Returning to the observations, using that, for zenith angles whose
$\cos \theta_z >-.6 $  $P_{\mbox{Super K}}=.3$ \cite{re2b}
we derive our second
constraint (see Eq.(\ref{superk})) on the mixing angles,
\bea
 s_2^2 c_2^2 c_3^4 = .15
\label{dos}
\eea

\section{Solar neutrinos}

As is well known, the four experiments set up to detect solar
neutrinos found significant deviations from the fluxes expected on the 
basis of the standard solar model
\cite{re1,solar}. The observed suppression is often
interpreted in terms of matter induced resonant oscillation, the 
MSW effect, by which, on its way through the sun's interior, the
initial electron neutrinos are turned
into other neutrino species to which the detectors are blind.

With neutrino mass differences as large as those we are
considering in this work, the MSW mechanism is not expected to be relevant. 
Therefore, the survival probability of a solar electron neutrino is given by
\bea
P_{\mbox{solar}} &= &1 -4 U_{13}^2  U_{11}^2 
\sin^2 \left[ \frac{(\Delta m^2 + \Delta M^2) L}{4 E} \right] \nn \\
&& -4 U_{13}^2 U_{12}^2 
\sin^2 \left[ \frac{\Delta M^2 L}{4 E} \right] 
-4 U_{12}^2 U_{11}^2 
\sin^2 \left[ \frac{\Delta m^2  L}{4 E} \right]
\label{solar}
\eea
Again, when the corresponding flight length and energy of the neutrinos
detected on Earth allow us to average the sine squared terms
to one half, this probability is given by
\bea
P_{\mbox{solar}}= 1 - 2 c_3^2 \left( s_3^2 + s_1^2 c_1^2 c_3^2 \right)
\eea

Although the Chlorine, the Gallium and the Kamioka experiments, probing
different parts of the solar neutrino spectrum, seem to find
somewhat different suppression factors we summarize their combined
information by taking $P_{\mbox{solar}}=.5$ \cite{solar}. 
This gives us a third
equation fixing the mixing angles,
\bea
1 - 2 c_3^2 \left( s_3^2 + s_1^2 c_1^2 c_3^2 \right) =.5
\label{tres}
\eea

\section{The mixing pattern and its consequences}

Solving Eqs. (\ref{uno}), (\ref{dos}) and (\ref{tres}), we 
obtain a neutrino mixing pattern that is compatible
at the same time with the LSND, the solar and the atmospheric neutrino
anomalies. In fact, we find at first four solutions that we show 
in Table \ref{table1}, two of which can be ruled out on
phenomenological grounds, (see below).

\begin{table}[hbt]
\begin{center}
\begin{tabular}{||c|c|c|c||} \hline
    $\theta_1$       & $\theta_2$ & $\theta_3$ & CKM\\ \hline
35.5     & 27.3 & 13.1 & 
$\pmatrix{ .793 & .566 & .226 \cr
          -.601 &.662 & .447 \cr
           .103 &-.490 &.865 }$ \\ \hline
40.4     & 64.2 & 6.5 & 
$\pmatrix{ .757 & .644 & .113 \cr
          -.360 &.266 & .894 \cr
           .546 &-.718 &.432 }$
\\ \hline
49.6  & 64.2 & 6.5 &
$\pmatrix{ .644 & .757 & .113 \cr
          -.398 &.204 & .894 \cr
           .654 &-.621 &.433 }$
\\ \hline
54.5  & 27.3  & 13.1 & 
$\pmatrix{ .566 & .792 & .226 \cr
          -.783 &.432 & .447 \cr
           .257 &-.430 &.865 }$
\\ \hline
\end{tabular}
\end{center}
\caption{Mixing patterns}
\label{table1}
\end{table} 

At this point we stress that we restricted ourselves
to the central values of the experimental data. 
Clearly, experimental error bars in the inputs will
reflect themselves in uncertainties of the resulting parameters. 
Here, however, our aim was primarily
to show that the central values can account essentially for all
observations.

It is also important to notice that, like other, previous attempts to 
explain neutrino phenomena with oscillations, ours is subject to the
resolution of possible inconsistencies between experiments.
For example, we have assumed that the LSND results is an oscillation
effect rather than a background one. However, one should keep in mind
that while  the LSND collaboration has announced additional evidence 
\cite{lsnd}
KARMEN has not confirmed their result \cite{kar} \footnote{For our
assumed parameters KARMEN should have yielded 
two or three real events on top of other three
background ones}.

The next question is how to distinguish between the four solutions we
give in Table 1.
The answer is clear: try with reactor experiments.
Eq. (\ref{solar}) which we have used for solar neutrinos is also valid
for reactor experiments. Although these experiments are rather insensitive
to the smaller of the assumed differencies 
$\Delta m^2 \left|_{\mbox{\small max }}= 10^{-3} eV^2 \right.$,
 they should observe
oscillations  with $\Delta M^2 =.3 eV^2$. For values of $E$ and $L$ 
that allow an averaging of the sine-squared term, we have
\bea
P_{\mbox{reactor}}= 1 -  2 s_3^2 c_3^2 
\eea
We therefore predict $ P_{\mbox{reactor}}=.90$ in cases I and IV and
$ P_{\mbox{reactor}}=.97$ (cases II and III).
This must be compared with the Bugey result of
$.99 \pm .01 (\mbox{stat}) \pm .005 (\mbox{sys})$ and the CHOOZ result of
$.98 \pm .04 (\mbox{stat}) \pm .004 (\mbox{sys})$.
With regard to the experimental errors we can state that all four
solutions are in reasonable agreement with the reactor data.

Up to this point all experimental phenomena which are believed to be
due to neutrino oscillations, can be understood by the mixing matrices
of cases I and IV and by squared mass differences in the range
$10^{-4} eV^2 < \Delta m^2 < 10^{-3 } eV^2$ and $\Delta M^2 = .3 eV^2
$.
For reasons to be discussed below, the second and third solution are
ruled out on phenomenological grounds.
Ours sets of parameters seem to be in reasonable agreement with all the 
features of neutrino observations, including the recently announced
zenith angle modulation of the Super Kamiokande atmospheric neutrino data
for upward  going events ($\cos\theta_z <-.6$).

Now, provided the atmospheric muon neutrinos are produced at almost 
twice the rate of the electron neutrinos, the double ratio for these
upward going events can be taken to be
\bea
R &=& \frac{ \left( \frac{N_\mu}{N_e} \right)_{\mbox{observed}}}
{ \left( \frac{N_\mu}{N_e} \right)_{\mbox{no oscillation}}} \nn \\
\nn \\
&= & \frac{1}{2}\; 
\left(\frac{ 2 P_{\mu \mu} + P_{e\mu}} {P_{ee} + 2 P_{\mu e}}
\right)
\eea
Assuming $L$ to be sufficiently large so that all oscillation
factors average to one half we obtain  $R_{large\mbox{ L}}$= .44 (case
I) 
and .53 (case IV)
to be compared with the experimental value $R_{exp}=.41 \pm .04$ 
at the largest zenith angles.
For cases II and III this double ratio reaches values as large as one and
it is for that reason that we should discard them.

As a more detailed test of our assumptions we have also calculated the
ratio (19) as a function of $L/E$, i.e. without averaging the
oscillation factors. Figs.~1 and 2 show the comparison with the zenith
angle dependence as reported by the Super-Kamiokande group
\cite{re2b}, for the two limiting values of $\Delta m^2$,
Eq.~1. Fig.~1 corresponds to solution I, Fig.~2 to solution IV of
Table~1. The figures show that the observed variation of $R$ with
zenith angle, or oscillation length, is well described in either case;
the comparison does not allow, as yet, to discriminate between cases I
and IV.

On top of that, the three neutrino mixing scheme predicts 
 sizeable oscillations
of electron to tau neutrinos that should be observed by the long 
baseline neutrino experiments with proposed high intensity muon sources 
\cite{re13}.
The probability of such an oscillation in our model, is given by,
\bea
P &=& 4 s_3^2 s_2^2 c_3^2 
\sin^2 \left[ \frac{\Delta M^2 L}{4 E} \right] 
\eea
which is well above the expected sensitivity. The measurement of
electron to  tau oscillations in such long
baseline experiments
would test the three neutrino model.

When the mixing matrix is allowed to have a CP violating phase,
the CP violating off diagonal probability
differences are given by \cite{cpv}
\bea
\Delta P& =& P(\bar{\mu} \rightarrow \bar{e}) - P(\mu \rightarrow e) =
-\left(P(\bar{\mu} \rightarrow \bar{\tau}) - P(\mu \rightarrow \tau) \right) 
\nn\\
&=&P(\bar{e} \rightarrow \bar{\tau}) - P(e \rightarrow \tau) \\
&=& - 4 J \sin\delta \left[ \sin \Delta_{21} +
\sin \Delta_{32} + \sin\Delta_{31} \right] \nn
\eea
where $J$ is the Jarlskog invariant \cite{cecilia}, 
$J= s_1^2 c_1 s_2 c_2 s_3 c_3$ 
and
\bea
\Delta_{ij}= \Delta m^2_{ij} L/2 E
\eea
with $\delta$ being the CP violating phase in the mixing matrix. 
In this formula, the effects of the mixing are conveniently separated
from the effects of the masses, appearing in the $\Delta_{ij}$.

Within our model, $J \leq .07$ and $ \left[ \sin \Delta_{21} +
\sin \Delta_{32} + \sin\Delta_{31} \right] ·\simeq {\cal{O}} (1)$ for $L/E
= 730 \mbox{km}/10 GeV$ (relevant for MINOS) and also for $L/E
= 730 \mbox{km}/25 GeV$, (relevant for ICARUS). Hence,
$\Delta P$ can be as large as .07 (for these parameters, matter effects
are negligible \cite{sato}).

We have more implications. The neutrino mass spectrum implied
by our scenarios is given by,
\bea
m_1  &\simeq & m \nn \\
m_2 &\simeq &\sqrt{m^2 + \Delta m^2} \\
m_3 &\simeq &\sqrt{ m_2^2 + \Delta M^2} \nn
\eea
There are two limiting cases of interest assuming that the largest mass is 
in the $eV$ range.\\[4pt] 

\noindent (i) One is the hierarchical limit, in which $m$ is negligible.
Then $ m_2 \simeq 10^{-2} eV \gg m_1$ and $m_3 \simeq .6 eV$.\\
\noindent (ii) The second one is the nearly degenerate limit in which 
\bea
m_1 &\simeq & 1 eV \nn \\
m_2 &\simeq & 1 eV \\
m_3 &\simeq & 1.2 eV \nn
\eea
In this case the sum of the neutrino masses is $\sum_i m_i \simeq 3.2 eV$.
The cosmological density parameter associated with neutrinos becomes
\bea  
\Omega_\nu = .011 h^{-2} \sum_i m_i\, =\, .035 h^{-2} \,\simeq \,.2\, ,
\eea
for $h$ of about .5, and the amount of the neutrino dark matter
component along with cold dark matter makes for a viable and
testable scenario for mixed dark matter \cite{re8}.

If these neutrinos are Majorana particles, the effective mass
$< m_{\nu_e}> $  relevant in neutrinoless double beta decay analysis is
\bea
< m_{\nu_e}> = \sum_i U_{ei}^2  \, m_i
\eea
We find that in the case of hierarchical spectrum $< m_{\nu_e}> \simeq
.14 eV$ whereas in the degenerate case $< m_{\nu_e}> \simeq 1.5 eV$ 
(this could be somewhat smaller when CP phases are taken into account).
It is interesting that these values are in the range of what the double 
beta decay experiments can probe now and in the near future 
\cite{re9}.

\section{Conclusions}

This analysis shows that, when taking the experimental results at face
value, the evidence for neutrino oscillations is substantial.
Although none of the existing positive signals is beyond some 
criticism, their ensemble is quite compelling.

All the available experimental information from accelerator, atmospheric and
solar neutrinos is accounted for in the framework of an analysis
assuming mixing of just three flavours. The solutions that we found are not 
significantly contradicted by any existing result, all conflicting
evidence  being below the two sigma level.

In order to distinguish our analysis from the one with a single 
oscillation process
with a small $\Delta m^2$ requires precise  
measurements of the multi GeV, overhead ($\cos\theta_z \sim 1$)
events at Super Kamiokande. Our pattern of mixings  predicts, in $R \sim .69 
\; (85)  $ for case I (case IV) 
while the single oscillation one requires $R \sim 1$.

Perhaps  most important is our prediction that muon neutrino to electron
neutrino oscillations should be confirmed by an observation of an excess 
of high energy electrons such us upgoing shower events. An analysis of
the proposed, and perhaps already observed, mixing pattern in terms of
mass matrices of charged and neutral leptons along the lines of
ref.~\cite{tria} is in progress and will be published elsewhere.

\vspace{.5cm}

\begin{center}
{\bf Acknowledgements}
\end{center}

We are very grateful to J.Bernabeu and S. Petcov for enlightening
discussions. A post-doctoral fellowship
of the Graduiertenkolleg ``Elementarteilchenphysik bei 
mittleren und hohen Energien"
of the University of Mainz is also acknowledged.

\newpage

\begin{figure}[!ht]
  \begin{center}
  \epsfig{file=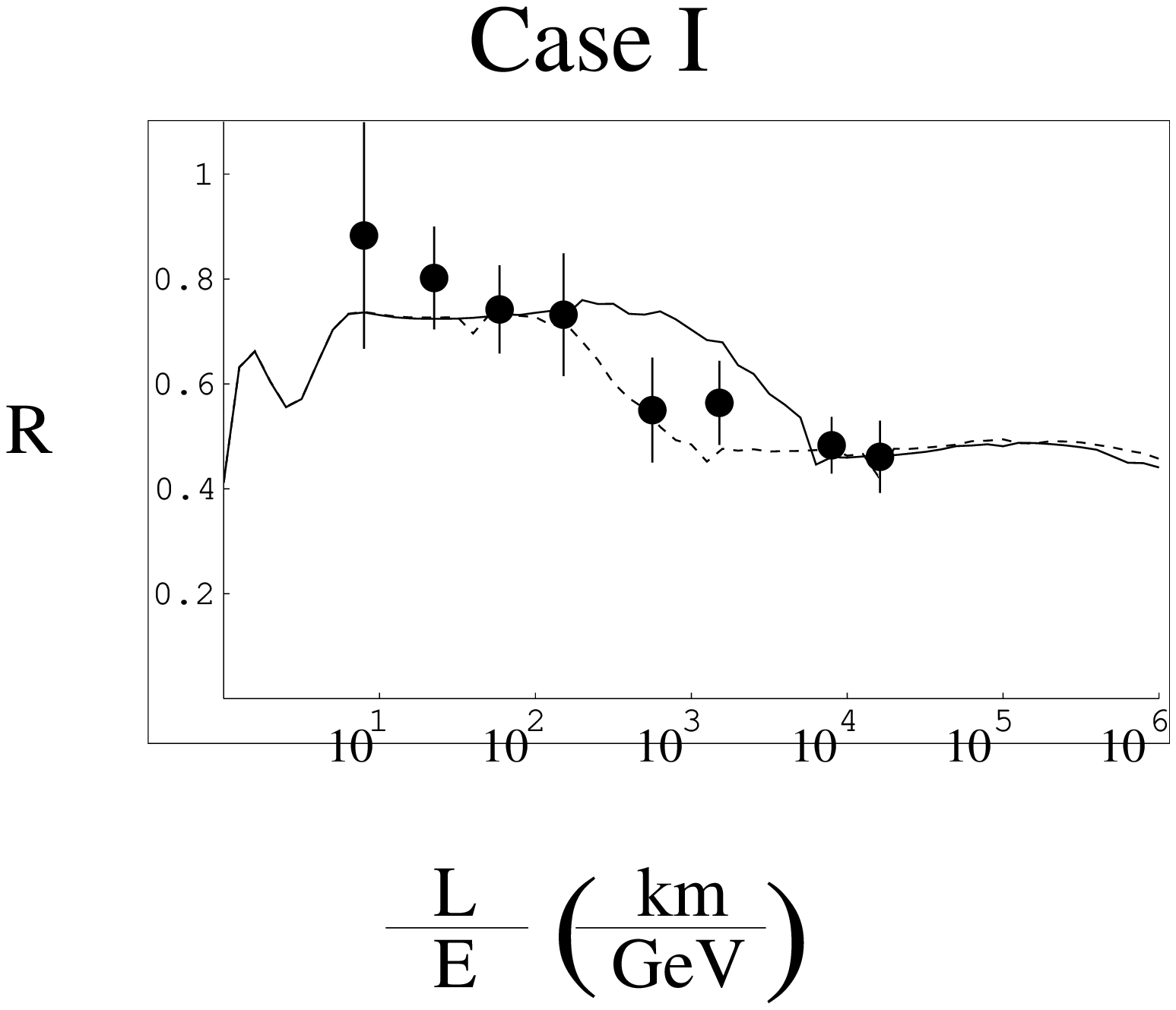,width=8cm,angle=270}
\caption{The ratio $R$, Eq.~(19), as a function of
  oscillation length over neutrino energy, $L/E$, in km/GeV,
  calculated for case I and compared to Super Kamiokande data. The
  solid line corresponds to the lower, the dashed line to the
  upper limit on $\Delta m^2$ of Eq.~1.} 
  \end{center}
\end{figure}
\begin{figure}[!ht]
  \begin{center}
  \epsfig{file=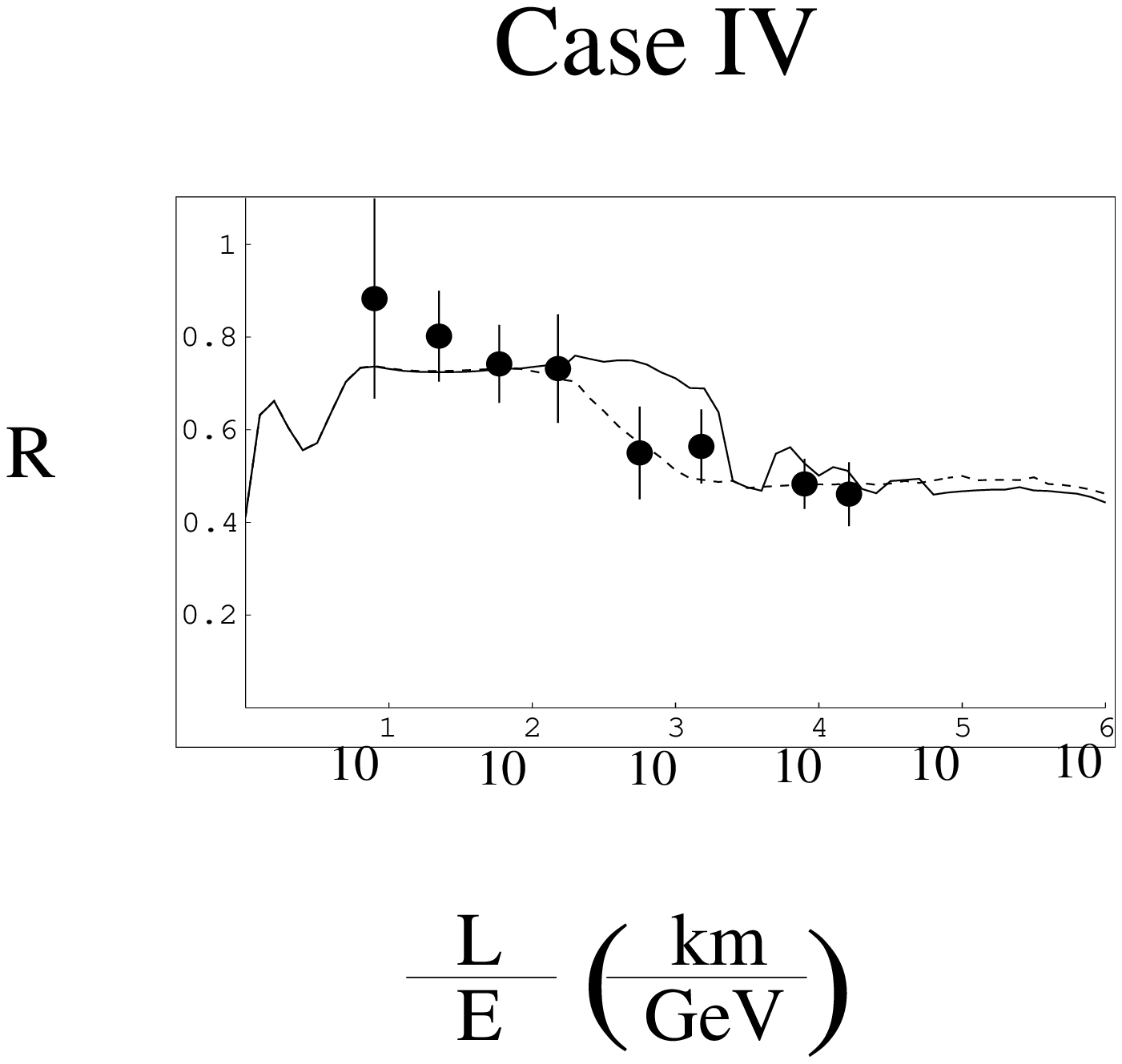,width=9cm,angle=270}
  \parbox{15cm}{\caption{Same as Fig.~1 for case IV.}}
  \end{center}
\end{figure}
\end{document}